\documentclass[a4paper,11pt]{article}
\pdfoutput=1 

\usepackage{jcappub} 

\usepackage[T1]{fontenc} 
\usepackage{subfig}

\def\be{\begin{equation}}
\def\ee{\end{equation}}
\def\barr{\begin{array}}
\def\earr{\end{array}}
\def\bea{\begin{eqnarray}}
\def\eea{\end{eqnarray}}
\def\bfig{\begin{figure}}
\def\efig{\end{figure}}
\def\calH{{\cal H}}
\def\eqn#1{eq.\ (\ref{#1})}
\def\fig#1{Fig.\ (\ref{#1})}
\def\lt{\left}
\def\rt{\right}
\newcommand{\nn}{\nonumber}

\title{Bounds on Neutrino Mass in Viscous Cosmology}

\author[a]{Sampurn Anand,}
\author[a]{Prakrut Chaubal,}
\author[a]{Arindam Mazumdar,}
\author[a]{Subhendra Mohanty,}
\author[a,b]{Priyank Parashari}

\affiliation[a]{Physical Research Laboratory, Ahmedabad, 380009, India}
\affiliation[b]{Indian Institute of Technology, Gandhinagar, 382355, India}

\emailAdd{sampurn@prl.res.in}
\emailAdd{prakrutchaubal@gmail.com}
\emailAdd{arindam@prl.res.in}
\emailAdd{mohanty@prl.res.in}
\emailAdd{parashari@prl.res.in}

\abstract{
Effective field theory of dark matter fluid on large scales predicts the presence of viscosity of the order of
$10^{-6} H_0 M_P^2$. It has been shown that this magnitude of viscosities can resolve the discordance 
between large scale structure observations and Planck CMB data in the $\sigma_8$-$\Omega_m^0$ and $H_0$-$\Omega_m^0$ 
parameters space. Massive neutrinos suppresses the matter power spectrum on the small length scales similar to the 
viscosities. We show that by including the effective viscosity, which arises from summing over non linear perturbations
at small length scales, severely constrains the cosmological bound on neutrino masses. Under a joint analysis of Planck CMB 
and different large scale observation data,  we find that  upper bound on
the sum of the neutrino masses at 2-$\sigma$ level, decreases from $\sum m_\nu \le 0.396\,$eV (normal hierarchy) and 
 $\sum m_\nu \le 0.378 \,$eV (inverted hierarchy) to $\sum m_\nu \le 0.267\,$eV (normal hierarchy) and $\sum m_\nu \le 0.146\,$eV 
 (inverted hierarchy) when the effective viscosities are included.
}


\begin{document}
\maketitle
\flushbottom

\section{Introduction}
\label{sec:intro}
In the past few decades, neutrino oscillation experiments have established the fact that neutrinos are
massive. Since the oscillation probability depends on the difference of squared masses of neutrinos, 
these experiments are insensitive to their absolute mass. Moreover, these experiments can not determine the 
mass ordering of three generation of neutrinos either. Thus, determination of the absolute masses of 
neutrinos and their ordering remains to be an open issue in particle physics and need to be settled. Apart
from particle physics experiments, cosmological observations can also be used to extract these informations
about neutrinos since they play a crucial role in the background evolution as well as formation of 
structures in the universe. 

In the recent past, some discordances between cosmic microwave 
background (CMB) and large scale structure (LSS) surveys have been reported. In particular, the value of 
$\sigma_8$, the r.m.s. fluctuation of density perturbations at 8 $h^{-1}$Mpc scale and $H_0$ the value
of Hubble parameter observed today, inferred from CMB and LSS observations are not in agreement with 
each other\cite{Vikhlinin:2008ym,Macaulay:2013swa,Battye:2014qga,MacCrann:2014wfa,Aylor:2017haa,Raveri:2015maa,Lin:2017ikq}.
It was argued that the mismatch between the Planck CMB observations and LSS surveys could be
a signature of nonzero neutrino masses and hence bound on sum of neutrino masses was obtained~\cite{Battye:2013xqa}. 
However, these attempts could not resolve the discordance in the above mentioned cosmological observations
simultaneously. Furthermore, those mismatches are shown to be resolved in a better way by using 
effective viscous description of cold dark matter (CDM) on large scales. In this framework,
the above mentioned discrepancies can be lifted simultaneously. With the success of effective 
viscous framework in resolving the discordance between CMB and LSS experiments, we will move on to include
massive neutrinos as well. In this paper we will demonstrate that in the effective viscous framework, we 
can indeed constrain the absolute masses of the neutrinos in a stringent way and differentiate the
mass ordering of neutrinos to some extent.

Ascribing the aforementioned discordance to some exotic physics, several attempts have been made to address 
the issue. For instance, the interaction between dark matter and dark energy~\cite{Pourtsidou:2016ico, Salvatelli:2014zta,Yang:2014gza} as well as dark matter and dark radiation~\cite{Ko:2016uft,Ko:2016fcd,Ko:2017uyb} was studied to resolve this tension to some extent.
Similarly, in other attempts neutrino 
sector has been modified~\cite{Wyman:2013lza,Battye:2013xqa,Riemer-Sorensen:2013jsa}. 
Massive sterile neutrino in the system was reported to reduce the tension in $\sigma_8$-$\Omega_m^0$ plane but it fails to do that in $H_0-\Omega_m^0$~\cite{Wyman:2013lza,Battye:2013xqa}. Interestingly, in ref~\cite{Anand:2017wsj} it has been shown that viscosities in CDM on large scales solves both $\sigma_8$ and 
$H_0$ tension simultaneously. We, therefore, use the viscous dark matter description to the study of 
neutrinos.

In order to demonstrate the tension between LSS and Planck CMB observation quantitatively, we use the 
following data sets. To study the LSS sector we used Planck SZ survey~\cite{Ade:2013lmv}, Planck lensing survey~\cite{Ade:2013tyw}, 
Baryon Acoustic Oscillation data from BOSS~\cite{Anderson:2013zyy,Font-Ribera:2013wce},
South Pole Telescope (SPT)~\cite{Schaffer:2011mz,vanEngelen:2012va} and CFHTLens~\cite{Kilbinger:2012qz,Heymans:2013fya}.
This combined data set will be referred as LSS data in this paper. However, by Planck data we mean 
only Planck CMB observation\cite{Ade:2015xua}.

This paper is structured as follows: We start with a brief description of tensions between CMB and LSS 
observations in section \ref{sec:cosmo-obs}. Further, we describe the massive neutrinos in 
section \ref{sec:nu-cosmo} followed by effective viscous frame in section \ref{sec:eff-visc}. 
After describing the massive neutrinos and viscous cosmology, we move on to the linear perturbation theory in
section \ref{sec:ptbn}. This section is divided further into three subsections. In
\ref{sec:ptbn_cdm} we give the perturbation equations for viscous CDM while in \ref{sec:ptbn_nu} we 
provide the set of equations for massive neutrinos used in CLASS code and in \ref{sec:effect_pk} we 
discuss the effect of them on matter power spectrum. In section \ref{sec:massive-nu-vdm}, we demonstrate that 
the inclusion of massive neutrinos can ease the tension between cosmological observations but do not solve 
them completely. However, the effective viscous framework solves the tension completely.  After 
demonstrating the success of viscous CDM model we perform similar Markov Chain Monte Carlo (MCMC) 
analyses in section \ref{sec:nu-m0-param} 
to constrain the neutrino parameter space and finally conclude in section \ref{sec:conc}.
\section{Tensions in cosmological observations}{}
\label{sec:cosmo-obs}
Large scale structure observations through lensing and Sunyaev-Zeldovich (SZ) effect have been consistently
reported some deficiency in the number of clusters from the expected value obtained  from CMB-fitted  parameters.
The problem can be described in the following way.

\paragraph{$\sigma_8$-$\Omega_m^0$ tension :}
Lensing observations estimates the power spectrum of the lensing potential ($C^{\phi\phi}_\ell$). This
$C^{\phi\phi}_\ell$ depends on two parameters, amplitude of primordial perturbations, $A_s$, and
the scale corresponding to the matter-radiation equality, $k_{\rm eq}$~\cite{Pan:2014xua,Ade:2015zua}. 
The amplitude of the matter power
spectrum $P(k)$ increases if the value of $A_s$ goes up. However, when $k_{\rm eq}$ changes, $P(k)$ gets shifted. 
Therefore there exist a degeneracy in $A_s$ and $k_{\rm eq}$ for a fixed value of $P(k)$ at certain $k$. 
This effect gets manifested in $C^{\phi\phi}_\ell$ too. $\sigma_8$ is propositional to $A_s$ and depends $\Omega_m$
through the growth factor. Similarly $k_{\rm eq}\equiv a_{\rm eq} H_{\rm eq}\propto \Omega_m h^2$. Therefore the
degeneracy in $A_s$ and $k_{\rm eq}$, when converted in terms of standard cosmological parameters,
shows up as a degeneracy in $\sigma_8$ and $\Omega_m$. 

In the case of SZ surveys what is measured is the number of clusters with given mass in a given volume along the
line of sight. This number is also a combination of $\sigma_8$ and the growth factor which is characterized by 
$\Omega_m$. Therefore most of the LSS observations  report their likelihood in 
$\sigma_8$-$\Omega_m^0$ plane as
\bea
\sigma_8\left(\Omega_m^0\over \Omega_{m \rm,\, ref}\right)^{\alpha} = {\rm const}\, .
\eea
The values of $\alpha$ and $\Omega_{m \rm,\, ref}$ are fitted so that the above combination remain
independent of $\Omega_m$. Therefore $\alpha$ and $\Omega_{m \rm,\, ref}$ changes for different observations.

\begin{figure}[!t]
\begin{center}
\hspace{-1.2cm}
\subfloat[\label{fig:sig8-om-planckLSS}]{
\includegraphics[width=3.17in,height=2.79in,angle=0]{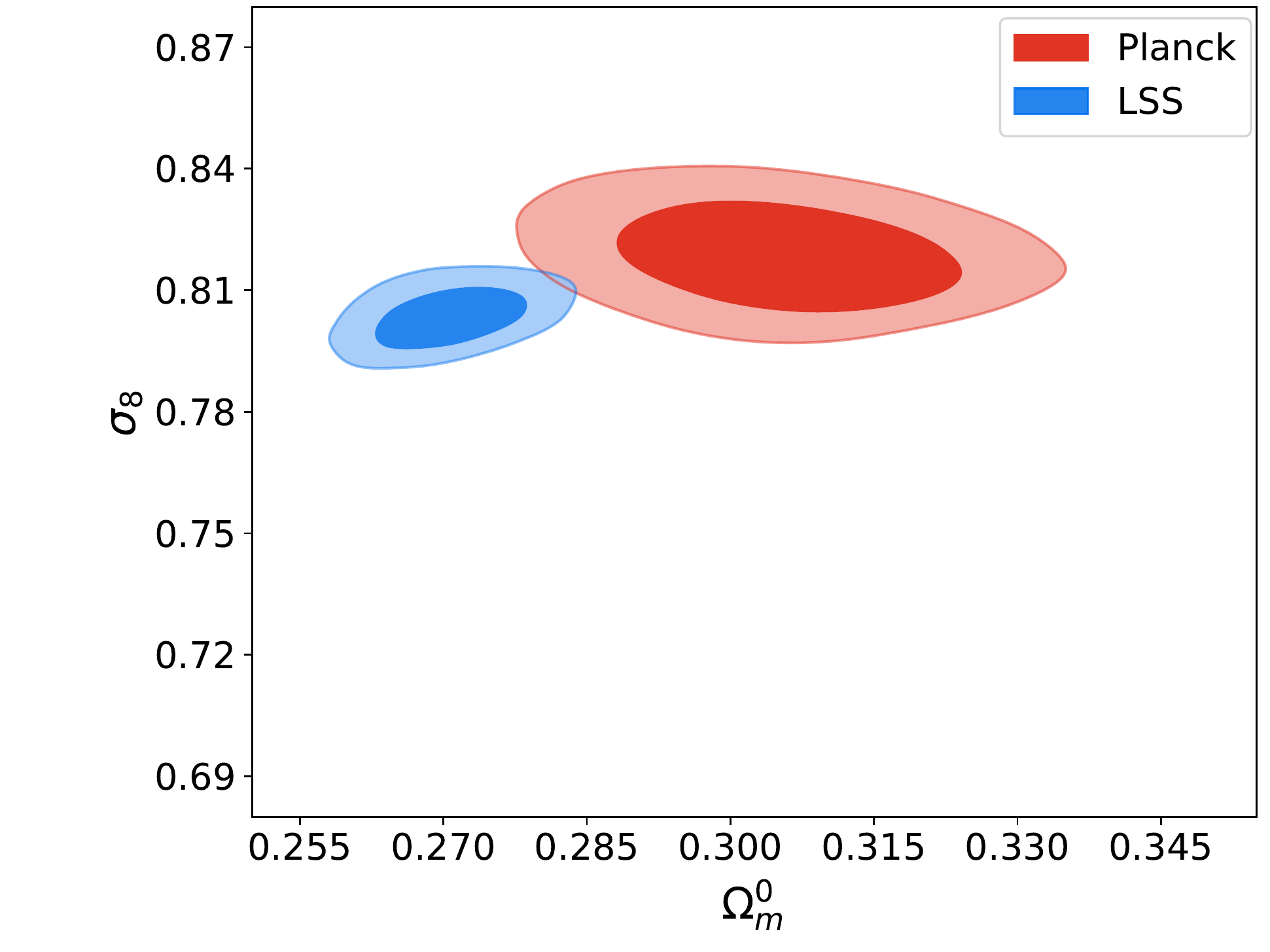} }
\subfloat[\label{fig:H0-om-planckLSS}]{
\includegraphics[width=3.03in,height=2.82in,angle=0]{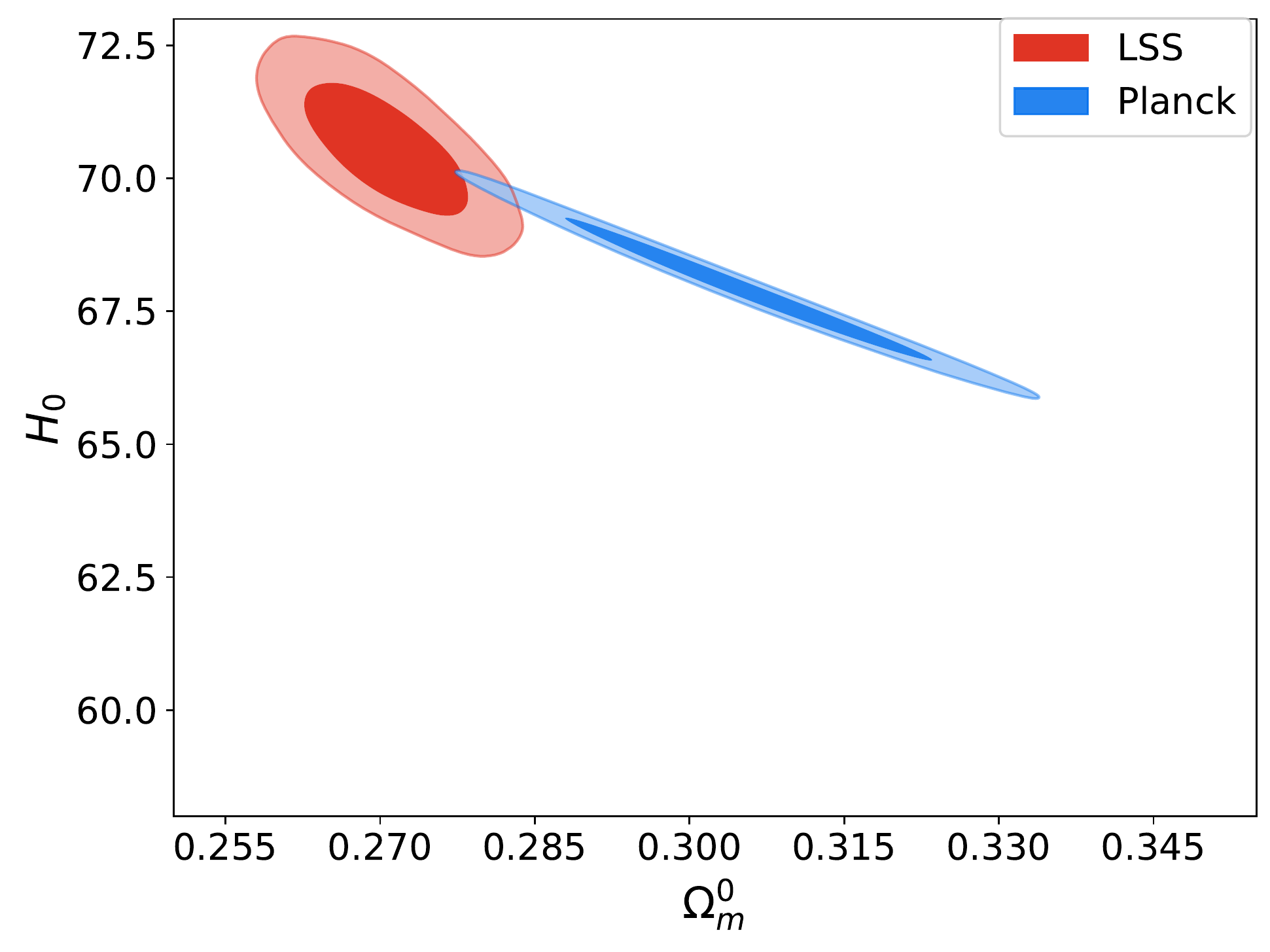} }
\caption{a) There exists a tension in allowed parameter space of $\sigma_8$-$\Omega_m^0$ from LSS observations
and Planck CMB observations. b) Similarly the mismatch in the allowed values of $\Omega_m^0$ manifests a mismatch
in $H_0-\Omega_m^0$ plane.}
\end{center}
\end{figure}

However, the bestfit value of $\Omega_m^0$ and $A_s$ obtained from the CMB experiments gives a value of 
$\sigma_8$ from the theoretically predicted matter power spectrum using $\Lambda$CDM cosmology.
This value does not match with the $\sigma_8$-$\Omega_m^0$ degeneracy direction at 2-$\sigma$ level 
(see \fig{fig:sig8-om-planckLSS}). 
Joint analyses by combining different LSS experiments remove the degeneracy in the $\sigma_8$-$\Omega_m^0$. 
But still a mismatch between the allowed region of $\sigma_8$-$\Omega_m^0$ from Planck CMB and LSS experiments 
persists. 

\paragraph{$H_0$-$\Omega_m^0$ tension :}
Measurement of $H_0$ is done in an indirect way in CMB and LSS observations. The scales of 
baryon acoustic oscillation (BAO) at last scattering surface, $\theta_{\rm MC}$ is actually observed in CMB. 
Similarly acoustic oscillation in the matter power spectrum is also observed by LSS surveys like SDSS.
Since, the co-moving acoustic oscillation scale is taken as the standard ruler in cosmology we can determine the
co-moving distance from BAO~\cite{Bassett:2009mm}. The comoving distance at a particular $z$ is 
\bea
\chi(z) = \int_0^z {dz'\over H(z')}\, ,
\eea
where,
\bea
H(z)^2 = H_0^2 (\Omega_m^0(1+z)^3 + \Omega_\Lambda)\, .
\eea
Therefore BAO observations can provide the value of $H_0$ when the value of $\Omega_m^0$ is supplied. 
Joint analyses of LSS experiments give some best-fit value of $\Omega_m^0$ rather than a large range. 
This $\Omega_m^0$ is little less than the $\Omega_m^0$ obtained from Planck CMB observations, which 
makes the value of $H_0$ derived from LSS joint analysis little higher than that derived from Planck CMB
observation as seen in \fig{fig:H0-om-planckLSS}.

This mismatch in $\sigma_8$-$\Omega_m^0$ and $H_0-\Omega_m$ is a problem which can be solved simultaneously if 
one can reduce $\sigma_8$ without adding some extra matter component in the theory. In literature this problem
has been addressed in many different ways as discussed in the sec.~\ref{sec:intro}. Adding massive neutrinos 
is one of those attempts. But massive neutrinos increase $\Omega_m^0$ while reducing $\sigma_8$. 
On the other hand effective viscous description of dark matter on large scales does not increase $\Omega_m^0$
for reducing $\sigma_8$. A comparison between these two different ways of easing tension will be discussed later
in sec.~\ref{sec:massive-nu-vdm}. In next three sections we will discuss the available understanding of
mass of neutrinos, effective viscosity of CDM and the modifications in standard perturbation theory 
due to the inclusion of these two non-standard parameters.

\section{Massive neutrino in cosmology} 
\label{sec:nu-cosmo}
Several experiments with solar, atmospheric and reactor neutrinos have established 
the phenomenon of neutrino oscillation, caused by non-zero neutrino mass and neutrino 
mixing, beyond doubt. Theoretically, neutrino oscillation can be explained successfully by 
assuming that the neutrino flavor eigenstates namely, $\nu_e,~\nu_\mu, ~\nu_\tau$  are linear 
superposition of mass eigenstates $\nu_1,~\nu_2,~\nu_3$. The fundamental parameters which
characterize the neutrino mixing are: three mixing angles, $\theta_{12}, \theta_{13}, 
\theta_{23}$, three neutrino masses, $m_1, m_2, m_3$ and the CP violating phase 
$\delta_{\rm cp}$. For the purpose of this paper we will call the mass of the lightest neutrino
as $m_0$.

Global analyses of neutrino data provide us the neutrino oscillation parameters namely,
$\Delta m_{21}^2, \theta_{12}, |\Delta m_{31}^2| (|\Delta m_{32}^2|), \theta_{13}, 
\theta_{23}$ with high precision. Here,
\bea
\Delta m_{ij}^2 = m_i^2 - m_j^2\, .
\eea
However, the existing data is blind towards the sign of
$\Delta m_{31}^2 $ or $\Delta m_{32}^2 $ and hence give rise to two types of neutrino mass 
spectrum. Conventionally, these two spectra are given as follows:
\begin{itemize}
\item the normal hierarchy: \\
\be
m_1(m_0) < m_2 < m_3,~~~ \Delta m_{31}^2 > 0,~~~ \Delta m_{21}^2 > 0,
\ee
\be\label{norm-mass}
m_2 = \sqrt{m_0^2 + \Delta m_{21}^2},~~~~ m_3 = \sqrt{m_0^2 + \Delta m_{31}^2}, 
\ee
\item the inverted hierarchy: \\
\be
m_3(m_0) < m_1 < m_2,~~~ \Delta m_{32}^2 < 0,~~~ \Delta m_{21}^2 > 0,
\ee
\be\label{inv-mass}
m_2 = \sqrt{m_0^2 + \Delta m_{23}^2}\, ,~~~~ m_1 = \sqrt{m_0^2 + \Delta m_{23}^2 - \Delta m_{21}^2}\, , 
\ee
\end{itemize}
The solar neutrino data tells us that the best fit values of $\Delta m_{21}^2 = 7.37 \times 10 ^{-5} {\rm eV}^2$. The 
values of $ \Delta m_{32}^2$ and $ \Delta m_{31}^2$ is obtained from the global fit of the atmospheric neutrino oscillation
data. In the case of normal hierarchy the bestfit value of $|\Delta m_{31}^2| $ is $2.54 \times 10 ^{-3} {\rm eV}^2$  and 
for inverted hierarchy $|\Delta m_{32}^2| $ is $ 2.42 \times 10 ^{-3} {\rm eV}^2$ ~\cite{Capozzi:2016rtj,Patrignani:2016xqp}.
We use these bestfit values in calculating the neutrino masses in CLASS code~\cite{Blas:2011rf}. However, while providing
2-$\sigma$ upper bound on the the sum of the neutrino masses we will be using the upper 2-$\sigma$ value of these quantities
from ref~\cite{Capozzi:2016rtj}.

Massive neutrinos have important property that they are relativistic in the early universe
and contributes to the radiation density. However, in the late time, when they turn non-relativistic they contribute
to the total matter density.
We would also like to highlight that
the collisionless nature of the neutrinos, after it become non-relativistic, allow them to 
free-stream on scales $k > k_{\rm fs}$, where $k_{\rm fs}$ is wavenumber corresponding to the scale of 
neutrino free-streaming. Hence it will wash out the perturbations on length scales smaller than 
the characteristic scale $k_{\rm fs}$. This leads to suppression of power on small scales in the matter 
power spectrum and modifies the gravitational lensing potential sourced by cosmological 
structures which in turn modifies the shape of CMB anisotropy. 

The number density of the neutrinos is very high and even small mass in neutrino can 
change the matter or radiation density of the universe. The radiation density corresponding to
the neutrinos in early universe, after it decouples from the electron-positron plasma, is given 
by~\cite{Lesgourgues:2006nd,lesgourgues2013neutrino}
\bea
\rho_{\nu} = \left[{7\over 8}\left(4\over 11\right)^{4/3}N_{\rm eff}\right]\rho_\gamma\, ,
\eea
where, $\rho_\gamma$ is the photon density and $N_{\rm eff}$ is the effective number of 
relativistic neutrinos at early times. The value of $N_{\rm eff}$ has been estimated to be 3.046~\cite{Mangano:2005cc}.
The slight change of the value of $N_{\rm eff}$ from exact 3 is attributed to the effect of spectral
distortion of the neutrinos just after the decoupling. At the late time, when these neutrinos turns into non-relativistic 
species, their energy density fraction in the present universe becomes
\bea
\Omega_\nu = \frac{\sum_i m_i}{\rm eV}{1\over 93.1 h^2}\, .
\eea
This $\Omega_\nu$ contributes to the total matter density fraction $\Omega_m$.  

\section{Viscous cold dark matter}
\label{sec:eff-visc}
Viscosities in the cold dark matter can be sourced by two different ways. When viscosities are 
generated due to self interaction between the dark matter particles, they are called
fundamental viscosities. There is a second type of viscosity which is known as effective viscosity. 
This kind of viscosities are expected to be generated on large scales as the integrated effect of the 
back-reaction of small scales non-linearities.

Bounds on the ratio of cross section to the mass of dark matter particles are extracted from bullet cluster 
observation~\cite{Kahlhoefer:2013dca}.
Fundamental viscosities can be calculated from these bounds on self interactions between the dark matter 
particles if the underlying quantum field theory of dark matter is well formulated~\cite{Kubo:1957mj} or an 
energy distribution function of the dark matter particles is known~\cite{Gavin:1985ph, Atreya:2017pny}. 
Nevertheless, the fundamental viscosities are expected to be small compared to the effective viscosities~\cite{Blas:2015tla}. 


Therefore we will concentrate on the second kind of the viscosities, known as effective viscosity. This kind 
of viscosities are expected to play its role 
in the late time of the universe. Linear perturbation theory works quite well for describing matter power spectrum
for a large range of scales. The reason behind this is the huge hierarchy between the virialization
scale and the Hubble scale. However, in late time when the non-linearities have started 
growing one can still compute an effective linearized perturbation theory on 
large cosmological scales taking into account the effect of small scales. It has been shown 
in the literature that the effect of these small scale dynamics when integrated out works 
as effective bulk and shear viscosities~\cite{Baumann:2010tm,Carrasco:2012cv}. Let us
decompose Einstein equation in background (with bar), linear (L) and non-linear (NL) parts.
\bea
\bar{G}_{\mu\nu} + G_{\mu\nu}^{\rm L} + G_{\mu\nu}^{\rm NL} = 8\pi G (\bar{T}_{\mu\nu}+{T}_{\mu\nu}^{\rm L}+{T}_{\mu\nu}^{\rm NL})\, ,
\eea
where, ${G}_{\mu\nu}$ is the Einstein tensor and $T^{\mu\nu}$ is the stress-energy tensor.
After equating $\bar{G}_{\mu\nu}$ and $8\pi G \bar{T}_{\mu\nu}$ we can write
\bea
G_{\mu\nu}^{\rm L} = 8\pi G \left({T}_{\mu\nu}^{\rm L}+{T}_{\mu\nu}^{\rm NL}-{G_{\mu\nu}^{\rm NL}\over 8\pi G}\right) 
   \equiv {T}^{v}_{\mu\nu} - \bar{T}^v_{\mu\nu}\, .
\eea
Here superscript $v$ stands for viscous fluid. In this formulation all the components of energy density like 
baryonic matter and CDM are expected to generate effective viscosities on large scales. For the sake of
simplicity we will consider viscosities only in dark matter fluid and treat baryonic matter 
as ideal fluid. We write the stress-energy tensor for non-ideal CDM fluid as~\cite{Weinberg:1972kfs}
  \be
  T^{\mu\nu}_{\rm cdm} = \rho_{\rm cdm}\,u^\mu\,u^\nu\, +\, (p + p_b)\,\Delta^{\mu\nu} +
  \pi^{\mu\nu}\, ,
  \label{eq:t-mu-nu-vf}
  \ee
where $\rho_{\rm cdm}$ is the energy density of the CDM and $p$ is the pressure which will be taken to be zero for further calculations.
Here, $u^\mu$ is fluid flow vector and 
  $p_b = -\zeta\, \nabla_\mu\,u^\mu$ is the bulk pressure with $\zeta$ being the coefficient of bulk viscosity.
$\pi^{\mu\nu}$ is the anisotropic stress tensor and has the following form
  \be
  \pi^{\mu\nu} =  -2\eta\,\sigma^{\mu\nu}
  = -2\eta\,
  \lt[\frac{1}{2}
    \lt(
 \Delta^{\mu\alpha}\nabla_\alpha u^\nu + \Delta^{\nu\alpha}\nabla_\alpha u^\mu
    \rt) - \frac{1}{3} \Delta^{\mu\nu}\lt(\nabla_\alpha u^\alpha \rt)
    \rt]\, ,
  \label{eq:shear-tensor}
  \ee
where $\eta$ is the coefficient of shear viscosity. 
For baryonic matter stress-energy tensor will same as \eqn{eq:t-mu-nu-vf} but the coefficients of
viscosities will vanish, since we are considering the baryonic matter to be ideal fluid.

The effects of viscosities are manifested in the cosmological parameters like, equation of state $w$ and 
sound speed $c_s^2$ and viscous sound speed $c_{\rm vis}^2$. 
\bea
w = -{3\zeta {\cal H}\over a \rho_{\rm cdm}}, ~~~~~~ c_s^2 = - {\zeta\theta\over a\rho_{\rm cdm}\delta}
~~~~~~~ c_{\rm vis}^2 = \left({4\over 3}\eta +\zeta\right){{\cal H}\over \rho_{\rm cdm}}
\eea
Here ${\cal H }$ is the comoving Hubble parameter, $\delta = {\delta\rho/ \rho}$ is the density
perturbation and $\theta = \nabla_i v^i$ is called the velocity perturbation. This velocity $v^i$
is the perturbation in spatial parts of fluid vector $u^{\mu}$.

For having an estimation of the viscous co-efficients we follow ref.~\cite{Blas:2015tla}.  
Let us take $k_m$ to be the scale beyond which non-linearities become important and we will
restrict our linear perturbation theory only up to that scale. Then, as assumed in ref.~\cite{Blas:2015tla}, 
the viscosity parameters vary in the following way,
\bea
c_s^2 &=& \alpha_s \left({\cal H}\over k_m\right)^2\, ,\nn\\
c_{\rm vis}^2 & = & \alpha_\nu (1+w)a \left({\cal H}\over k_m\right)^2\, .
\eea
If we assume $k_m$ to be order of 1 Mpc and take the value of all other quantities at to be of the present time,
then ${4\over 3}\eta +\zeta$ turns out to be of the order of $10^{-6}M_{P}^2 H_0 $. Here, $\alpha_s$ and $\alpha_\nu$
are kept to be order one for matching the results with N-body simulation~\cite{Carrasco:2012cv,Blas:2015tla}. Therefore, for $k_m \sim$ 1 Mpc,
$c_s^2$ also turn out to be of the order of $10^{-6}$.

As described in ref.~\cite{Anand:2017wsj} the effect of viscosities are mostly visible in the very late time of the 
growth of perturbations. In this period the value of the effective viscosities doesn't change that much; at least
the order remains same~\cite{Carrasco:2012cv}. Therefore we will consider constant viscosity for the time being in this paper.
Moreover, although there are two types of viscosities, bulk and shear, only a combination of them acts as a single 
parameter in the theory. So, we use only non-zero shear viscosity which mimics this combination. 
\section{Perturbation theory}\label{sec:ptbn}
After a brief discussion of viscous cold dark matter and massive neutrinos, we would like to
discuss the linear perturbation theory for these two components. Perturbation theory for these 
two components is dealt in two different ways. Although the massive neutrinos cannot travel freely
beyond the free streaming scale $k_{\rm fs}$, the mean-free path in between the neutrinos is infinity, since
neutrinos don't have self interaction in standard picture. 
This forbids us to treat neutrino as an fluid. 

Therefore we will treat CDM as viscous fluid and derive its perturbation theory by using conservation equation,
but for the neutrinos we will discuss the Boltzmann equations which describe the evolution of density and pressure 
perturbations of neutrinos.

\subsection{Perturbation theory with viscous cold dark matter}
\label{sec:ptbn_cdm}

We introduce perturbations in the CDM fluid in following way.
 \bea
\rho_{\rm cdm}(\tau, \vec x)& =& \rho_{\rm cdm}(\tau) + \delta\rho_{\rm cdm}(\tau, \vec x)\, .
\eea
In order to preserve homogeniety and isotropy, the 
background quantities are functions of time only whereas the perturbations are both space-time dependent. 
Perturbations in the FRW metric, taken in conformal Newtonian gauge, can be written as
\be
  ds^2 = a^2(\tau)\lt[-(1+ 2\,\psi(\tau, \vec x))\,d\tau^2 + (1-2\phi(\tau, \vec x))\,dx_i\,dx^i\rt]\, ,
  \label{eq:ptrb-metric}
  \ee
where $\psi(\tau, \vec x)$ and $\phi(\tau, \vec x)$ depends on space-time. This allows us to introduce 
first order perturbation in the fluid flow $u^\mu$ as
\bea
u^\mu = (1-\psi, v^i ) \, ,
\eea
while $u^\mu u_\mu = -1$ is preserved in first order limit.

The Friedmann equation for this system reads as
   \be
  \calH^2  = \lt(\frac{\dot a}{a}\rt)^2  = \frac{8\pi\,G}{3}\lt(\rho_{\rm m} + \Lambda\rt) a^2\, ,
  \label{eq:hubble}
  \ee
  where $\rho_{\rm m} = \rho_b + \rho_{\rm cdm} +\rho_\nu$ ,$i.e$ sum of baryonic matter density, CDM density and neutrino density.
  Here dot denotes the derivative with respect to conformal time $\tau$. The continuity equation of each 
  species in the background has the following form
  \be
  \dot \rho_i + 3\,\calH\,\lt(\rho_i + p_i\rt) = 0\, ,
  \label{eq:cont}
  \ee
In the case of neutrinos, when they are relativistic their pressure is $p_\nu = {1\over 3}\rho_\nu$. 
However, when neutrino temperature drops below its mass, its pressure drops down to small value and 
neutrinos become non-relativistic. Therefore, depending on the neutrino temperature at a particular time, different 
neutrino species with different mass might obey separate continuity equations.

The perturbed part of continuity equation for the cold dark matter, $T^{\mu\nu}_{~~;\mu} =0$, 
provides us two basis sets of equations, known as 
density perturbation equations and velocity perturbation equations~\cite{Anand:2017wsj}.  
\be 
  \dot\delta = -\lt(1 -\frac{\tilde\zeta\,a}{\Omega_{\rm cdm}\,\tilde\calH}\rt)
                   (\theta -3 \dot\phi)~
              + ~\lt(\frac{\tilde\zeta\, a}{\Omega_{\rm cdm}\, \tilde\calH}\rt)
   \theta ~ - ~\lt(\frac{3\,\,\calH\,\tilde\zeta\,a}{\Omega_{\rm cdm}}\rt)\,\delta \, 
\label{eq:delta}
\ee
and,
  \be 
  \dot\theta = -\calH\,\theta + k^2\psi 
        -\frac{k^2\,a\,\theta}{3\,\calH\,(\Omega_{\rm cdm}\,\tilde\calH-\tilde\zeta\, a)}
               \lt(\tilde\zeta + \frac{4\tilde\eta}{3}\rt)
        - 6\,\calH\,\theta\lt(1-{\Omega_{\rm cdm}\over4}\rt)
               \lt(\frac{\tilde\zeta\,a}{\Omega_{\rm cdm}\,\tilde\calH}\rt)\, ,
\label{eq:theta}
  \ee
 where $\tilde\eta = {8\pi G \eta\over{\cal H}_0}$ and $\tilde\zeta = {8\pi G \zeta\over{\cal H}_0}$ are the 
 dimensionless parameters constructed from the viscosity coefficients. After outlining the evolution equations for
 density and velocity perturbations in viscous CDM, we outline the same for massive neutrinos in the next sub-section.
\subsection{Perturbation theory with massive neutrinos}
\label{sec:ptbn_nu}
In this section, we will closely follow Ma \& Bertschinger~\cite{Ma:1995ey} to describe the perturbation equations for 
massive neutrinos. In terms of distribution function and the 4-momentum component, the energy momentum 
tensor is given by
\be
T_{\mu \nu} = \int dP_1 dP_2 dP_3 (-g)^{-1/2} {P_\mu P\nu \over P^0} f(x^i,P_j,\tau)\, ,
\ee
where $g$ and $P_\mu$ are the determinant of metric $g_{\mu \nu}$ and 4-momentum respectively and 
$f(x^i,P_j,\tau)$ is the phase space distribution function, which can be expressed as
\be
f(x^i,P_j,\tau) = f_0 (q)[1 + \Psi(x^i,P_j,\tau)] .
\ee
$f_0$ is the zeroth-order phase space distribution which is Fermi-Dirac for the case of neutrinos
and $\Psi$ is the perturbation in it. Massive 
neutrinos also obeys the collisionless Boltzmann equation. However, non-zero mass complicates the evolution
of distribution function. The unperturbed energy density and pressure of massive neutrinos are given by
\be
\bar \rho_h = 4\pi\,a^{-4}\int q^2dq \epsilon f_0(q), ~~~~ 
\bar P_h = {4\pi a^{-4}\over 3}\int q^2 dq {q^2\over \epsilon} f_0(q)
\ee
where $\epsilon = \epsilon(q,\tau) = \sqrt{q^2 + m_\nu^2 a^2}$. Unlike massless neutrinos, the $\epsilon$ depends
on both the momentum and time. As a result, we can not simplify our calculation by integrating out the momentum 
dependence in the distribution function. To proceed further, the perturbation $\Psi$ is expanded in a Legendre series
as
\be
\Psi(\vec k,\hat n, q,\tau) = \sum_{l =0}^{\infty} (-i)^l (2l +1)\,\Psi_l (\vec k, q,\tau) P_l(\hat k . \hat n)\, .
\ee
Using the above mentioned Legendre series expansion of $\Psi$, the perturbed energy density, pressure, 
energy flux and shear stress for the massive neutrinos in $k$ space can be given as
\bea
\delta \rho_h& =& 4 \pi a^{-4} \int q^2 dq \epsilon f_0(q) \Psi_0\, ,\nn\\ 
\delta P_h& =& {4 \pi \over 3} a^{-4} \int q^2 dq {q^2 \over \epsilon} f_0(q) \Psi_0\, ,\nn\\
(\bar{\rho}_h + \bar{P}_h) \theta_h & =& 4 \pi k a^{-4} \int q^2 dq q f_0(q) \Psi_1\, ,\nn\\ 
(\bar{\rho}_h + \bar{P}_h) \sigma_h & =& {8 \pi \over 3} a^{-4} \int q^2 dq {q^2 \over \epsilon} f_0(q) \Psi_2\, 
\label{eq:pert_nu_stress}.
\eea
Boltzmann equations for different moments of distribution function take the following forms in conformal
Newtonian gauge:
\bea\label{eq:boltz-nu}
\dot{\Psi}_0 &= & -{qk\over \epsilon} \Psi_1 - \dot{\phi} {d \ln f_0 \over d \ln q}\, ,\nn\\
\dot{\Psi}_1 &= & {qk\over 3 \epsilon}(\Psi_0 - 2\Psi_2) - {\epsilon k \over 3 q } \psi {d \ln f_0 \over d \ln q}\, ,\nn\\
\dot{\Psi}_l &= & {qk\over (2l + 1) \epsilon} [l \Psi_{l-1} - (l+1)\Psi_{l+1}]\, , \hspace{0.6cm} {\rm for,}\,\,  l\geq2.
\eea
The perturbations $\delta\rho_h$, $\delta P_h$, $\delta\theta_h$ and $\delta\sigma_h$ which comes from the 
energy momentum tensor are equated with the perturbed part of the Einstein equations. 
These perturbations in the energy momentum tensor are evaluated from \eqn{eq:pert_nu_stress} by 
numerically integrating the $\Psi$s over $q$. 
 
\subsection{Effect on matter power spectrum}\label{sec:effect_pk}
\begin{figure}[!tbp]
\begin{center}
\includegraphics[width=4.2in,height=3.2in,angle=0]{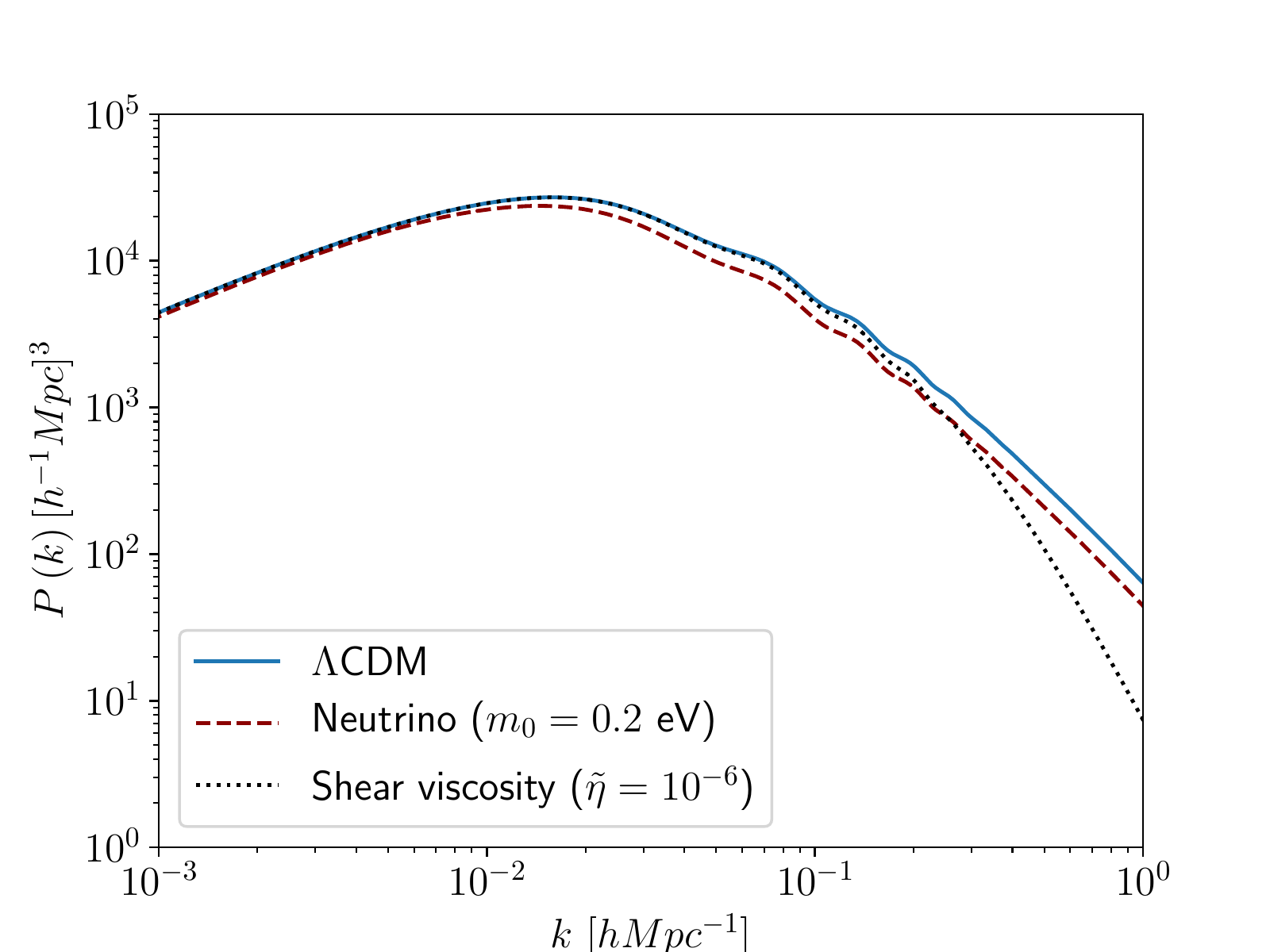}~~
\caption{Neutrino and viscous dark matter both suppress $P(k)$. However, while neutrinos effect are equal on all scales above 
$k_{\rm nr}$, effect of viscosity is strong only on the small scales. }\label{fig:pk}
\end{center}
\end{figure}
We have numerically solved the above mentioned Boltzmann hierarchy equations for neutrino, \eqn{eq:boltz-nu} and density-velocity
perturbation equations for viscous CDM, \eqn{eq:delta} and \eqn{eq:theta} using publicly available CLASS 
code~\cite{Blas:2011rf,Lesgourgues:2011rh}. The value
of lightest massive neutrino has been taken to be 0.2 eV  and masses of the other neutrinos have been taken as functions of it 
as described in sec.~\ref{sec:nu-cosmo}. Effect of the shear viscosity and massive neutrinos on matter power spectrum is plotted
in \fig{fig:pk}

The effect of viscous CDM and massive neutrinos have some similarities. The combination of $\zeta$ and $\eta$ in the third
term of the right hand side of \eqn{eq:theta} provides leading order contribution of the viscous effect. This term reduces
the velocity perturbations and feeds that to the \eqn{eq:delta} in which a reduced $\theta$ suppresses the 
growth of $\delta$. Therefore combination of these viscosities acts as a damping term in evolution of $\delta$. Since 
viscosity coefficients always come with $k$ in the perturbation equations their effect is larger in large $k$ values
(see \fig{fig:pk}). Since bulk and shear viscosity behaves in an indistinguishable way (see ref.~\cite{Anand:2017wsj})
we use only non-zero value of shear viscosity. 

In the case of neutrinos we see that Newtonian potentials $\psi$, created by the CDM perturbations, feeds in to the 
second moment of Boltzmann hierarchy equations, \eqn{eq:boltz-nu}. Iteratively it leaves its effect on all the $\Psi_l$s. Then the 
solution of Boltzmann equations changes the neutrino density perturbations through \eqn{eq:pert_nu_stress}. Power-spectrum 
generated from these massive neutrinos then adds up to the $P(k)$. However, this way of modifying 
neutrino density perturbations through the potentials of CDM is only possible on those scales which are larger than the 
free streaming scale corresponding to $k_{\rm fs}$. In general $k_{\rm fs}$ is a $z$ dependent quantity. It reaches
its minimum value $k_{\rm nr}$, where $k_{\rm nr}$ is the scale which reenters horizon at the time when neutrinos becomes
non-relativistic. This scale depends on the neutrino masses as~\cite{Lesgourgues:2014zoa}
\bea
k_{\rm nr} = 0.018\,\, (\Omega_m^0)^{1/2} \left(m_i\over 1\, {\rm eV}\right)^{1/2} h\,{\rm Mpc}^{-1}
\eea
For the particular mass $m_0 =0.2$eV, $k_{\rm nr}$ turns out to be $4.5\times 10^{-3}h\,{\rm Mpc}^{-1}$. Therefore, for 
the modes with $k$ smaller than this, neutrinos will fall in the potentials of CDM and for larger $k$s they will freely
steam without contributing in $P(k)$. That means in late time these neutrinos, on scales larger than $k_{\rm nr}$, will behave
as cold dark matter but won't form any structure. This effect is visible in the suppression of $P(k)$ above that scale in \fig{fig:pk}. 
\section{Massive neutrino and viscous dark matter as solutions}
\label{sec:massive-nu-vdm}

\begin{figure}[!tbp]
\begin{center}
 Massive neutrinos \hspace{4.5cm} Viscous CDM fluid \\
\hspace{-1.1cm}
\includegraphics[width=3.38in,height=2.78in,angle=0]{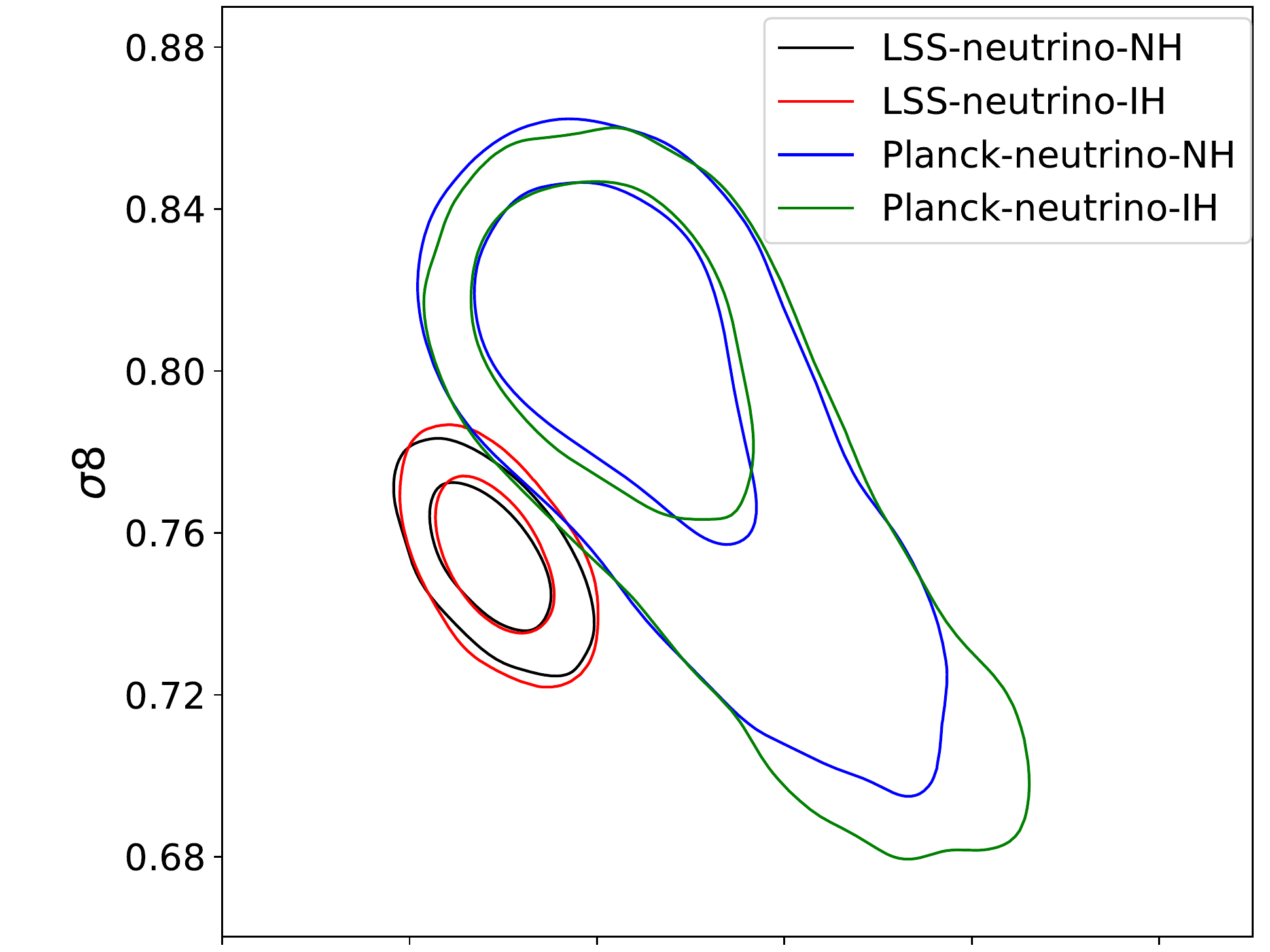}
\hspace{-0.32cm}
\includegraphics[width=2.95in,height=2.8in,angle=0]{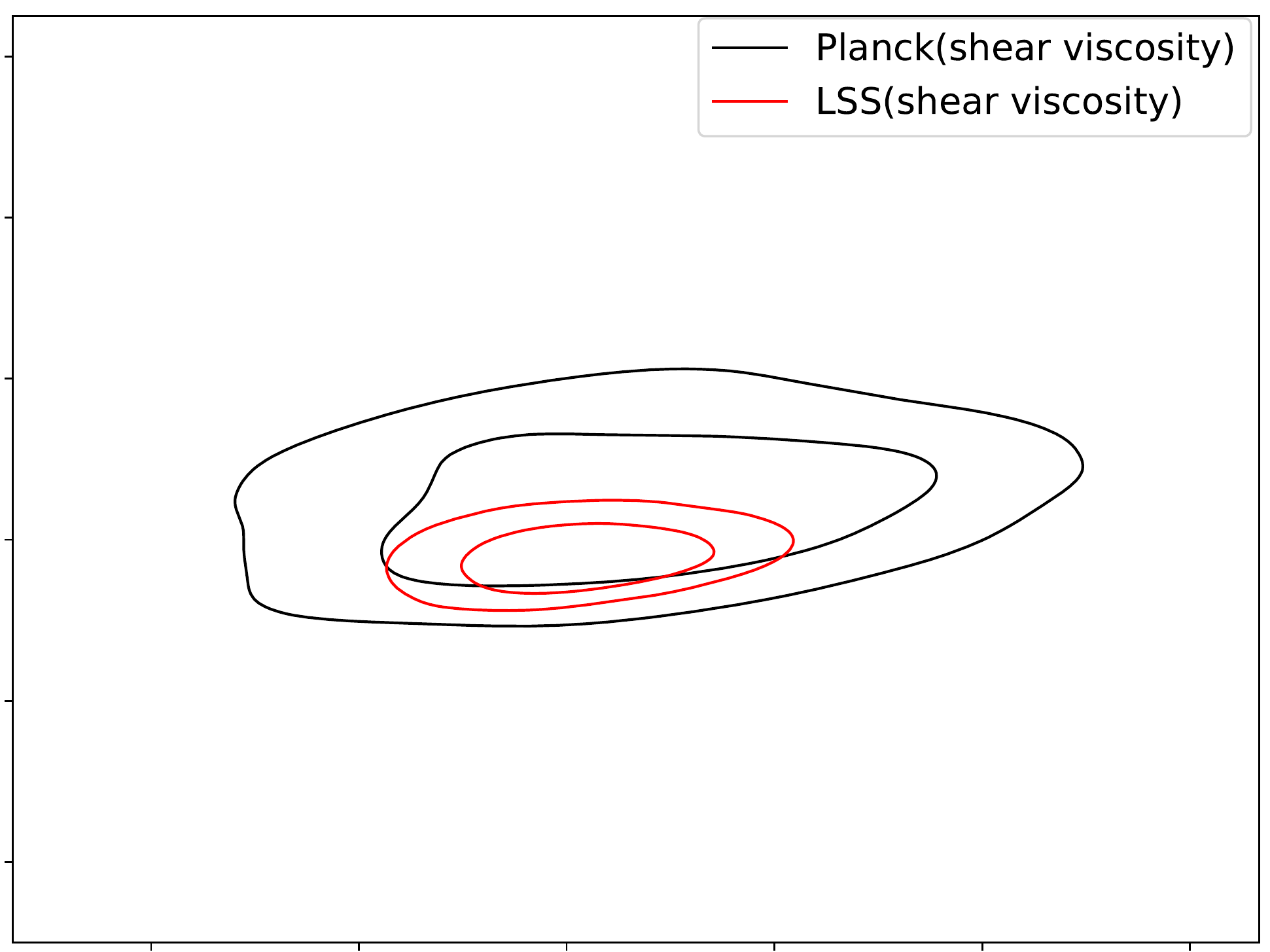}
\\
\vspace{-0.05cm}
\hspace{-1.1cm}
\includegraphics[width=3.34in,height=2.8in,angle=0]{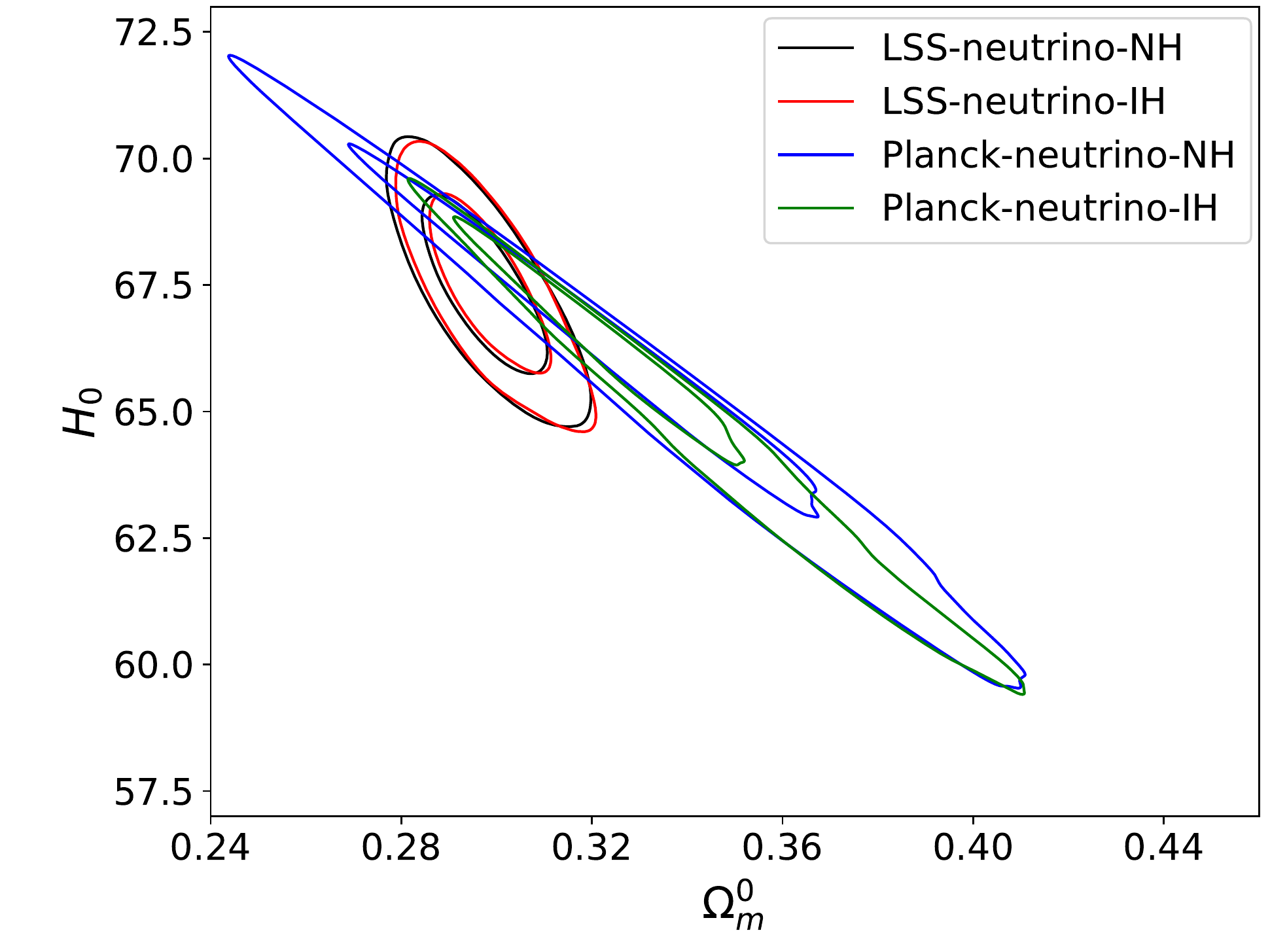}
\hspace{-0.24cm}
\includegraphics[width=2.95in,height=2.8in,angle=0]{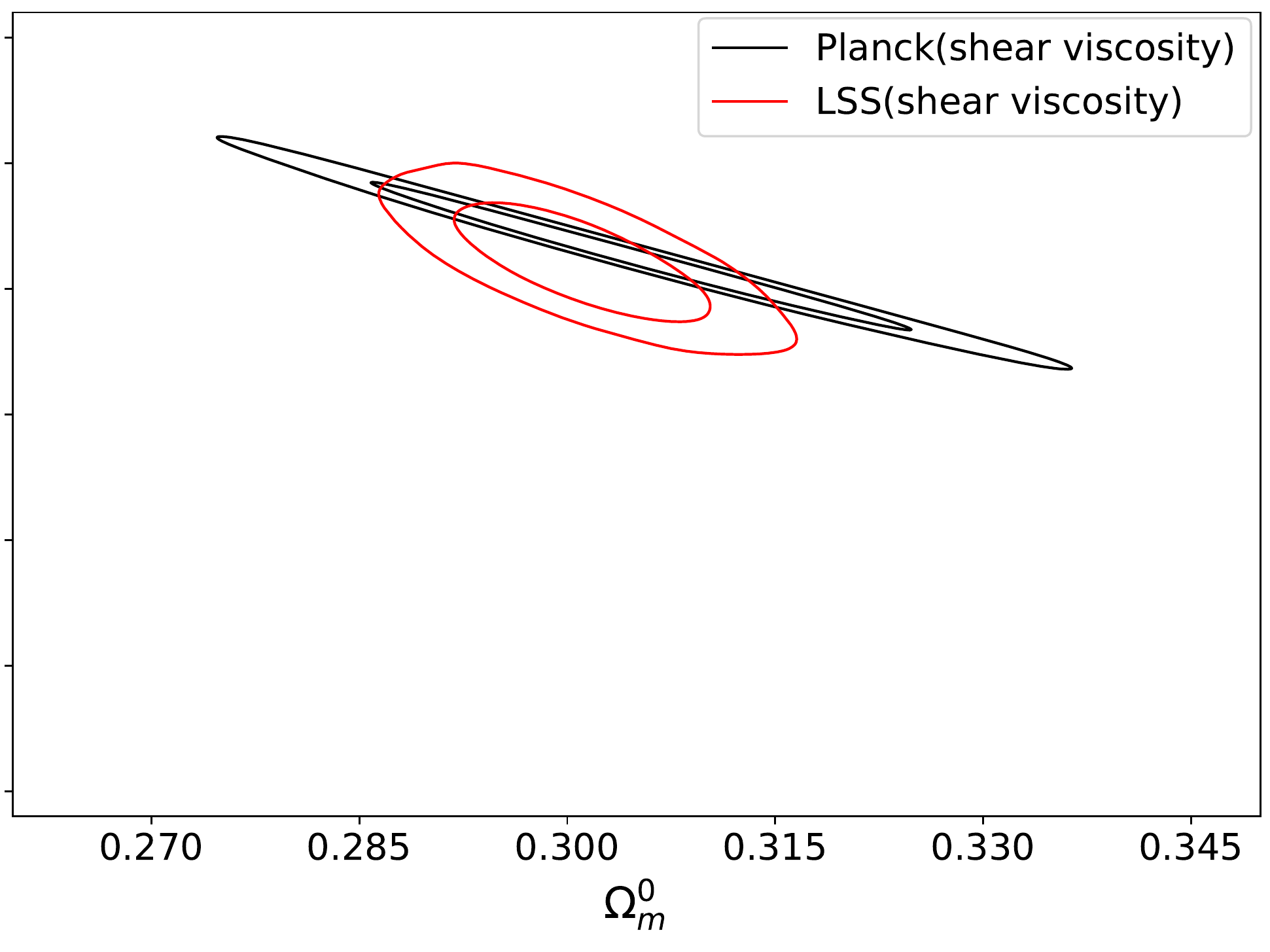} \\
(a)\hspace{6.8cm} (b) 
\caption{ (a) Inclusion of massive neutrinos helps to resolve the $H_0$ tension (lower panel), but
cannot solve the $\sigma_8$ tension simultaneously (upper panel). (b) Presence of any of the two 
viscosities resolves both the tensions simultaneously.}\label{joint-plots}
\end{center}
\end{figure}

We have already seen the effect of massive neutrinos and effective viscosity of CDM on matter power spectra (see \fig{fig:pk}).
Both the components suppresses $P(k)$ on small scales, though the effect of viscosity on the smaller scales are more
prominent than massive neutrinos.
Moreover, massive neutrinos increase the matter density at late time while
effective viscosity keep the matter density unaffected. To quantify their effect we proceed through the following steps.

First we run a MCMC analysis using MontePython~\cite{Audren:2012wb} of Planck $TT$ data with six cosmological parameters,
lightest neutrino mass $m_0$  
and $\sigma_8,\,  H_0$ as derived parameters. For normal hierarchy of neutrinos other two mass values are set
following \eqn{norm-mass} and similarly for inverted hierarchy following \eqn{inv-mass}.
Then the 2-$\sigma$ range obtained for two early universe parameters $A_s$ and $n_s$ are kept as bound
with flat prior on these two parameters in our next run with LSS data. In both the runs $m_0$ was varied
from range zero to 0.5 eV. 

Two derived parameters $\sigma_8$ and $H_0$ have been plotted against $\Omega_m^0$ in \fig{joint-plots}.
Since massive neutrino increases $\Omega_m^0$ while decreasing the $\sigma_8$ we can see that the allowed
regions for two set of experiments in $\sigma_8$-$\Omega_m^0$ plane moves side by side but never overlap. 
But in the $H_0-\Omega_m^0$ plane the varying neutrino mass enlarges the range of $H_0$ for LSS as well 
as Planck. This enlargement ultimately allows LSS value to be accommodated within the limits of Planck. 

For demonstrating the effect of viscosity we repeat the same steps, but this time keeping all the neutrinos massless
and fixing the effective shear viscosity parameter $\tilde\eta$ at $6\times 10^{-6}$. This value of shear 
viscosity has been reported to be the bestfit value in a joint Planck-LSS analysis in ref.~\cite{Anand:2017wsj}.
In the same paper it has been shown that bulk and shear viscosity plays indistinguishable role. Therefore we 
restricts our study to only one type of the viscosities. 

Effect of effective shear viscosity on $\sigma_8$-$\Omega_m^0$ and $H_0-\Omega_m$ plane has been demonstrated in 
\fig{joint-plots}. As discussed earlier shear viscosity damps the growth on small scales and reduces the 
value of $\sigma_8$ without adding any extra contribution to $\Omega_m^0$. Since there exists a degeneracy in
$\sigma_8$ and $\Omega_m^0$ for LSS data, as discussed in sec.~\ref{sec:cosmo-obs}, reduction in the value of $\sigma_8$
drives $\Omega_m^0$ to higher values. On the other hand, for CMB data the derived $\sigma_8$ gets reduced due to inclusion 
of viscosity and $\Omega_m^0$ remains unchanged. This helps to overlap the allowed regions in $\sigma_8$-$\Omega_m^0$
from two different sets of observations. Similarly, since $\Omega_m^0$ remains unchanged the derived values of 
$H_0$ from CMB data also does not get modified. However, the increase in the allowed values of $\Omega_m^0$ 
from LSS data reduces the derived values of  $H_0$. In this way effective viscosity helps to ease the tension 
at two different fronts, while neutrino can resolve only one of them.

These analyses motivate us to incorporate neutrino mass in the effective viscous CDM scenario. Therefore, we do 
a joint MCMC analysis with combined Planck and LSS data with extra two parameters namely mass of lightest 
neutrino $m_0$ and viscosity coefficient $\tilde\eta$. This provides us a bestfit value of $\tilde \eta$, which
is sufficient to ease the tension between LSS and Planck data, and a maximum allowed value of $m_0$.
We discuss this effect in details in the next section.


\section{Parameter space of neutrino mass}
\label{sec:nu-m0-param}
\begin{figure}[!tbp]
\begin{center}
\hspace{-1.1cm}
\subfloat[\label{fig:m0-bound-NH}Normal hierarchy]{
\includegraphics[width=3.2in,height=2.6in,angle=0]{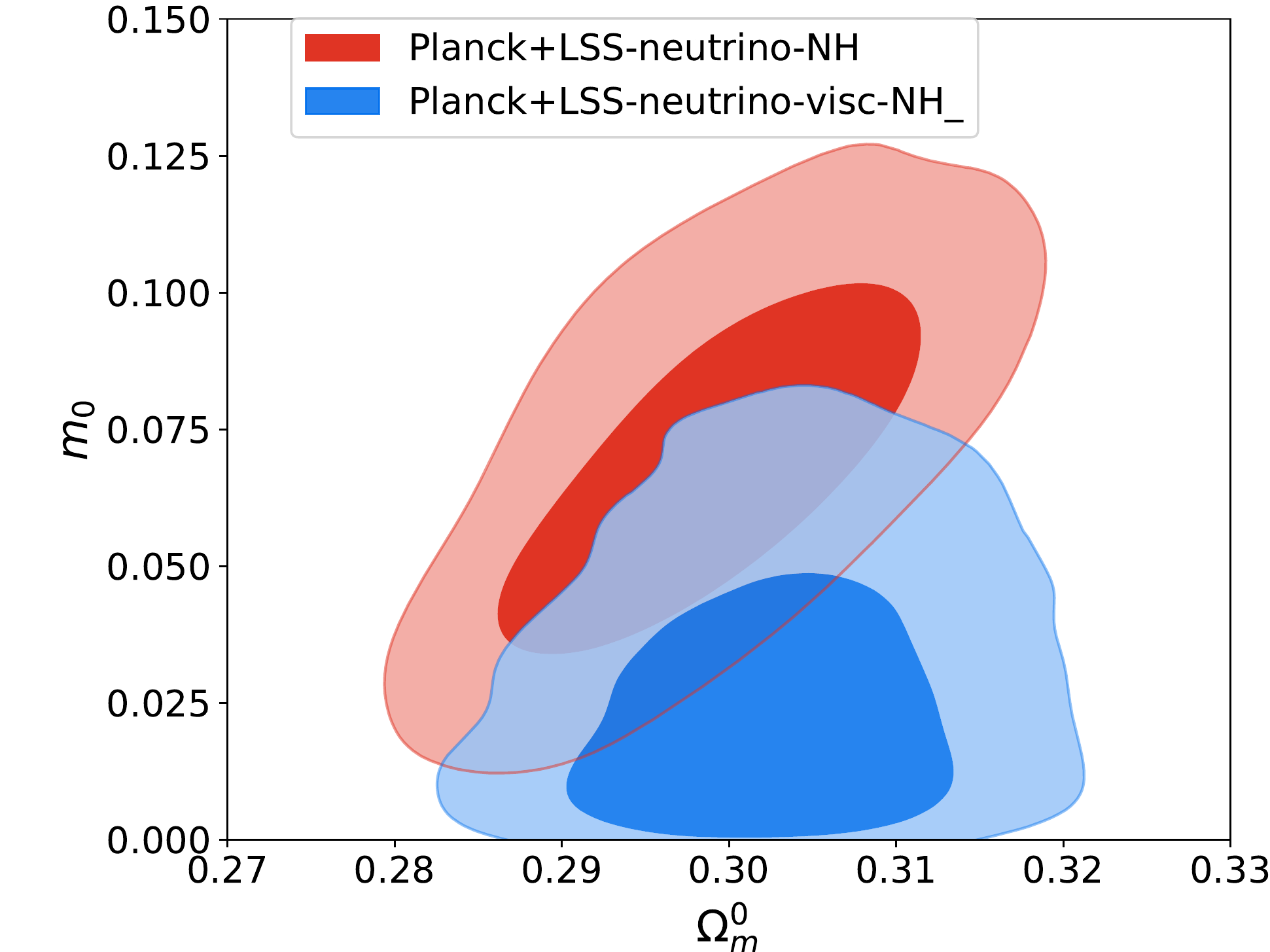}}~~
\hspace{-0.6cm}
\subfloat[\label{fig:m0-bound-IH}Inverted hierarchy]{
\includegraphics[width=3.2in,height=2.6in,angle=0]{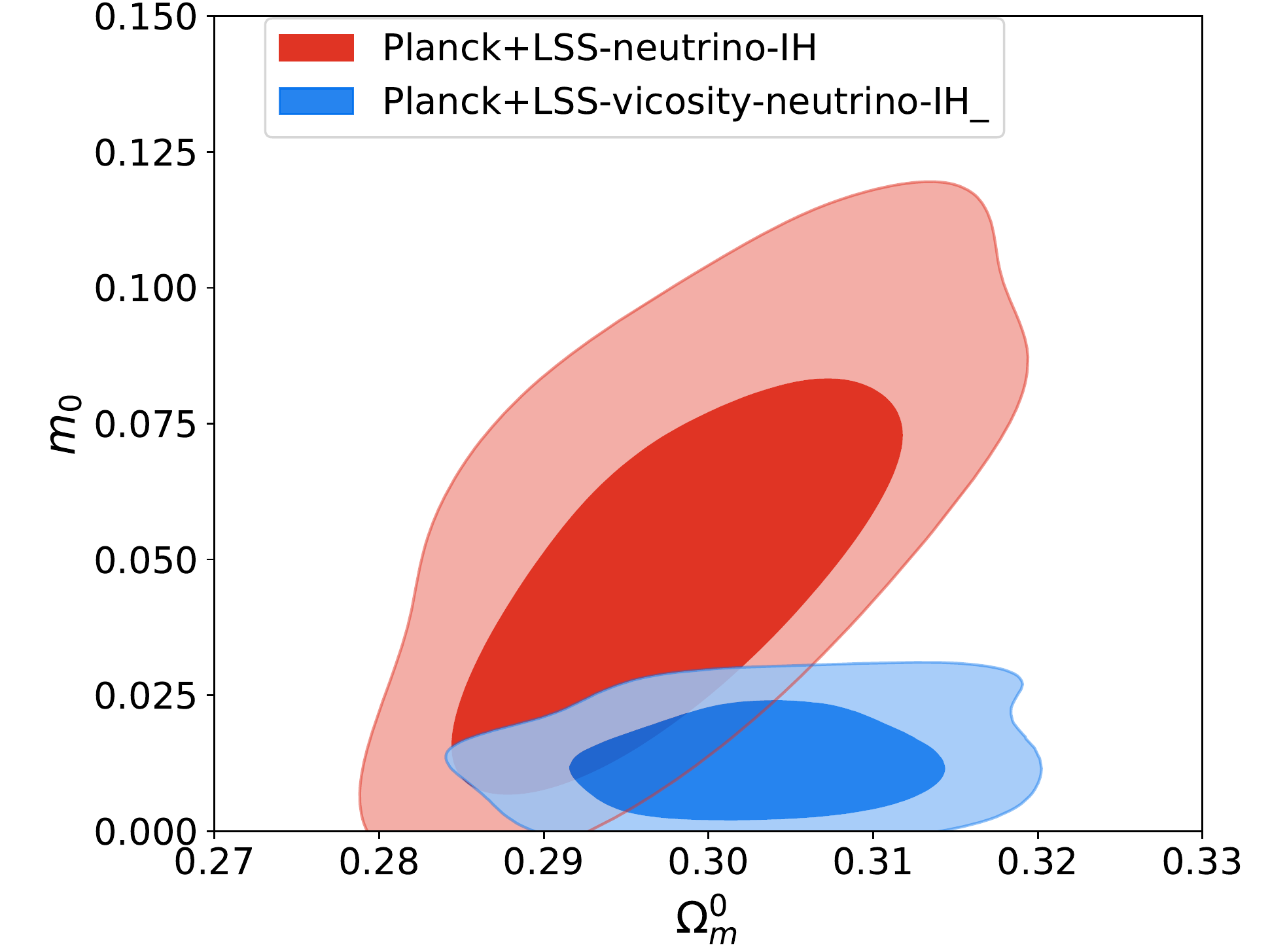}}
\caption{Planck CMB bound on the lowest neutrino mass $m_0$ drastically improves over inclusion of LSS 
data along with viscous cold dark matter.}
\end{center}
\end{figure}

\begin{figure}[!tbp]
\begin{center}
\hspace{-1.1cm}
\subfloat[\label{fig:m0-bound-NH}Normal hierarchy]{
\includegraphics[width=3.2in,height=2.6in,angle=0]{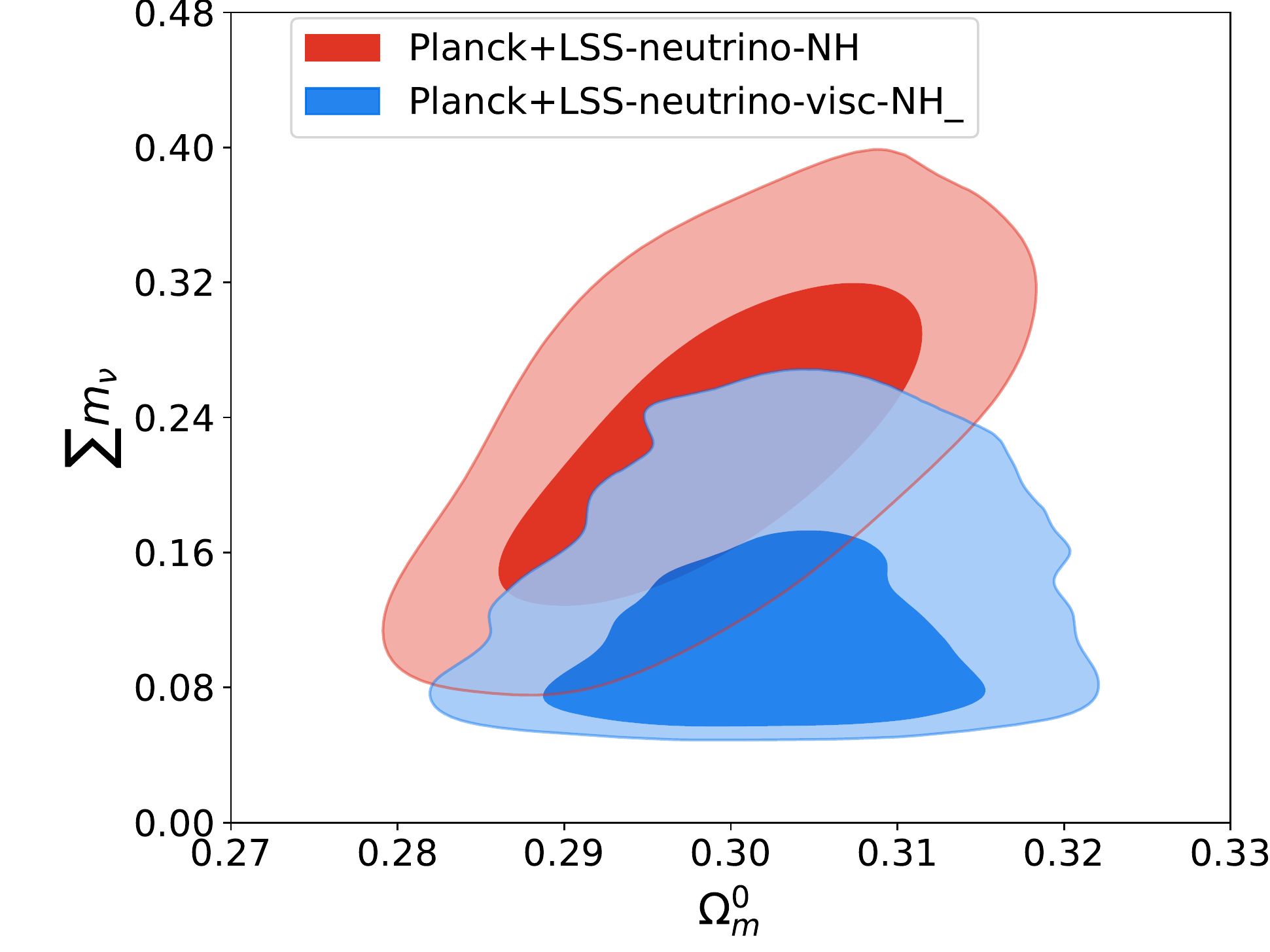}}~~
\hspace{-0.6cm}
\subfloat[\label{fig:m0-bound-IH}Inverted hierarchy]{
\includegraphics[width=3.2in,height=2.6in,angle=0]{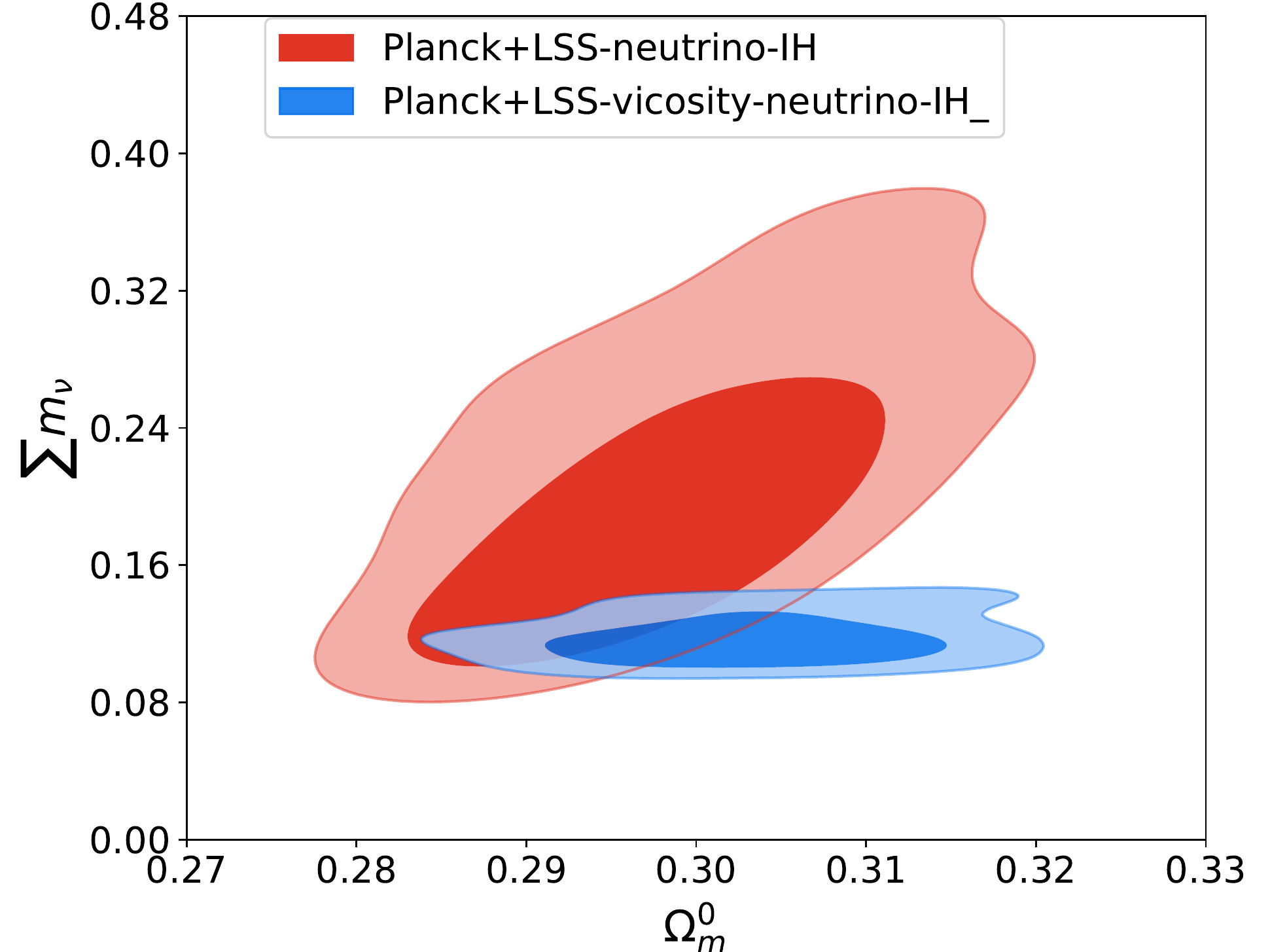}}
\caption{Planck CMB bound on the sum of neutrino masses $\sum m_\nu$  also drastically improves over inclusion 
of LSS data along with viscous cold dark matter.}
\end{center}
\end{figure}

As discussed earlier, inclusion of massive neutrinos in cosmology changes the shape 
of $P(k)$ which leads to reduction in the value of $\sigma_8$. Therefore it is expected
that better the measurement of $P(k)$ from large scale observations more stringent will
be the constraint on the neutrino mass.
However, precise measurements of $P(k)$ and CMB spectrum
in recent years has shown a discordance in between this two sets of observations which 
cannot be removed by inclusion of neutrino mass as it was thought earlier. Rather a better
description of cold dark matter fluid on large scale which includes the effect of small scale
nonlinearities as effective viscosity in linear regime, has been found sufficient to 
reconcile this discordance. However, neutrino oscillation experiments indicate that
neutrinos are massive. Consequently in the viscous fluid description it is expected
that we will get more stringent constraint on lowest neutrino mass.
In the joint MCMC analyses of Planck and LSS data in this frame work we found the maximum 
allowed value of $m_0$ at 2-$\sigma$ level to be 0.084 eV for normal hierarchy and 0.03 eV for inverted
hierarchy.
\\
\\
\begin{tabular}{ |p{3.5cm}||p{3.5cm}|p{3.5cm}|p{3.5cm}|  }
 \hline
 \multicolumn{4}{|c|}{ 2-$\sigma$ upper bounds on sum of the neutrino masses $\sum m_{\nu}$} \\
 \hline
 Data sets   	& Hierarchy & Without Viscosity & With viscosity\\
 \hline
 Planck + LSS    & NH  & 0.396 eV  & 0.267 eV  \\
 Planck + LSS    & IH  & 0.378 eV  & 0.146 eV  \\
  \hline
\end{tabular}
\vspace{0.5cm}

We have also performed a joint MCMC analysis of Planck and LSS data with only massive neutrinos (without viscosities) as 
it has been argued to provide the evidence of nonzero mass of lightest neutrino\cite{Wyman:2013lza,Battye:2013xqa}. We
found that in the case of normal hierarchy the 2-$\sigma$ lower bound of $m_0$ is 0.012 eV,
whereas for inverted hierarchy the 2-$\sigma$ bound accommodates zero. However, as discussed
above this preference of non-zero neutrino mass is nothing but an effect of not including 
effective viscosity in the analyses. From this same analysis we can also see that the upper bound on 
the $m_0$ goes up to 0.126 eV for normal and 0.119 eV for inverted hierarchy. That provides the 
sum of the neutrino masses $\sum m_\nu$ to be 0.396 eV and 0.378 eV respectively (taking all 
the other parameters at 2-$\sigma$ upper level).

We found that effective description of viscosity not only removes the notion of finding
non-zero mass of the lightest neutrino from cosmological observations but also tightens up the
bound on $m_0$ than that provided by the Planck+LSS joint analysis (see \fig{fig:m0-bound-NH} and \fig{fig:m0-bound-IH}).
The 2-$\sigma$ upper bound on $m_0$ for normal hierarchy results in the sum of the neutrino mass ($\sum m_\nu$)
to be 0.267 eV and for inverted hierarchy it goes up to 0.146 eV (taking all the other parameters at 2-$\sigma$ upper level). 
The value of $\sum m_\nu$ for NH is quite 
close to that of the Planck(TT+ BAO + HST) joint analyses in ref.~\cite{Ade:2015xua} with degenerate neutrinos.
However we find that the bound obtained for $\sum m_\nu$ in inverted case is much lower (see \fig{fig:msum}).

\subsection{Comparison with other experiments}
\begin{figure}[!tbp]
\begin{center}
\hspace{-0.5in}
\subfloat[\label{fig:msum}]{
\includegraphics[width=3.0in,height=2.6in,angle=0]{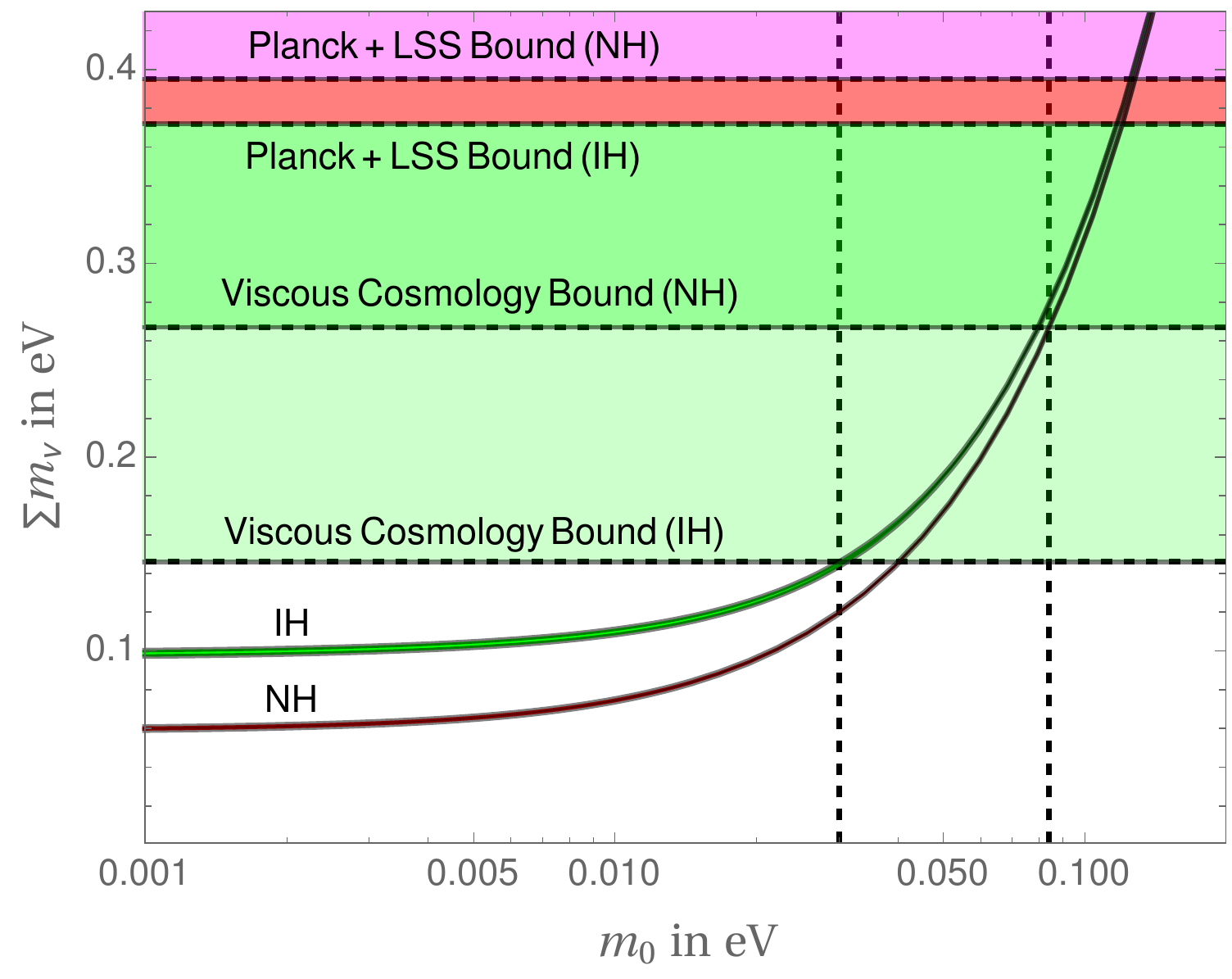}}~~
\subfloat[\label{fig:double-beta}]{
\includegraphics[width=3.0in,height=2.6in,angle=0]{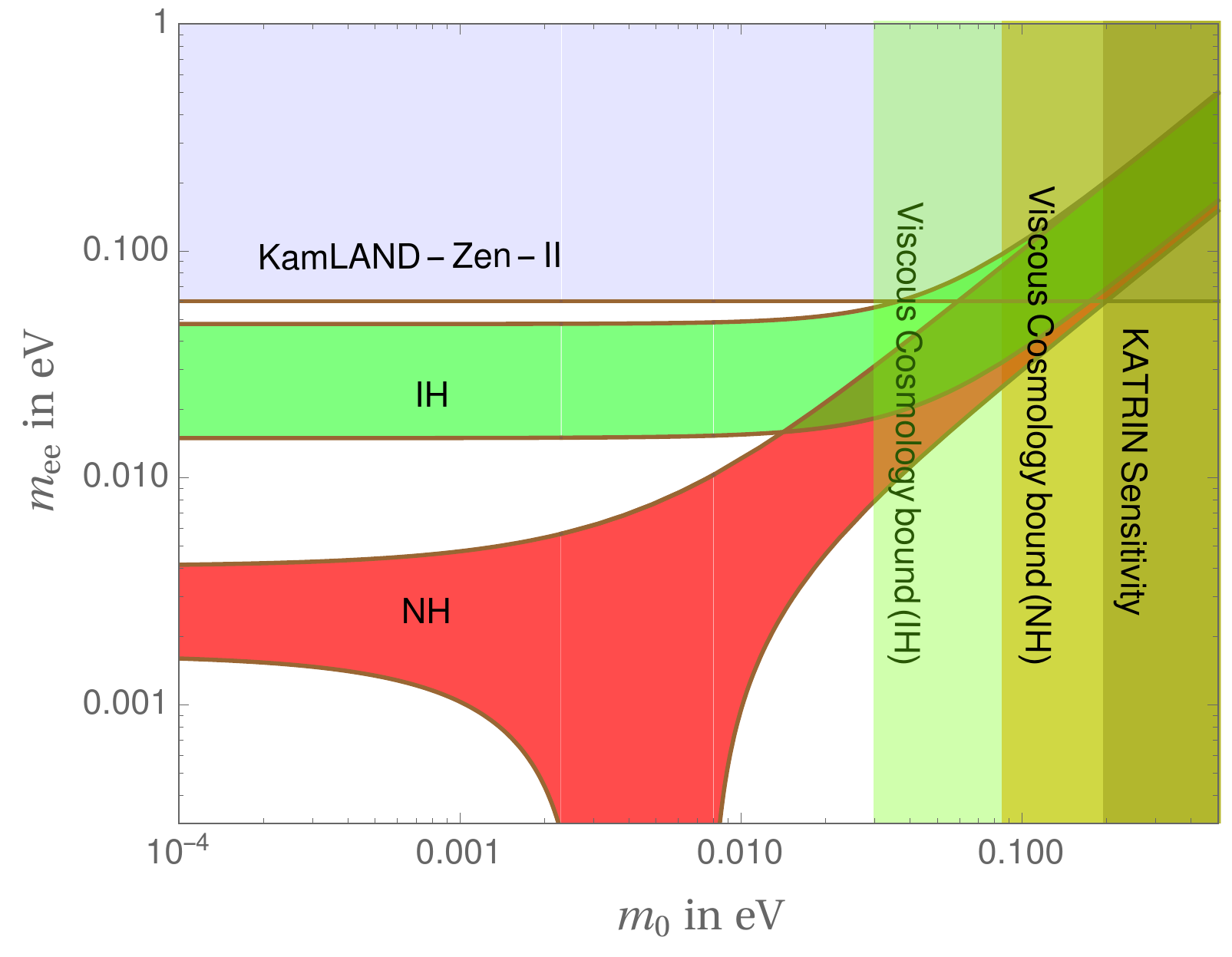}}
\caption{a) Sum of the neutrino masses as a function of $m_0$ are shown. The width of the of the function
corresponds to 2-$\sigma$ uncertainties in $\Delta m_{ij}^2$.
b) Mass of electron Majorana neutrino, $m_{ee}$ as a function of $m_0$ has been plotted where 
the values of other parameters are taken at 2-$\sigma$ level.}
\end{center}
\end{figure}

Except the cosmological observations there are two types of experiments which try to resolve the absolute
scale of mass of the neutrinos. The first type of experiments measure the $\beta$ decay spectrum and tries to
find the cutoff of the spectrum at the tail which is definite signature of neutrino mass. The expected 
sensitivity from this kind of experiments is best from KATRIN which is expected to probe $m_0$ up to 0.2 eV~\cite{Thummler:2010tt}.
Our bound on $m_0$ form normal hierarchy as well as inverted hierarchy goes below this sensitivity. Therefore
if viscous cosmology is the right solution to the tension between Planck and LSS observations, we are expected
to find no massive neutrino signal in KATRIN.

There are second types of experiments which try to observe the neutrinoless double beta decay which is 
not only a signature of neutrino mass but also the property of the neutrino, $i.e$ Dirac fermion or a Majorana fermion.
Probability of neutrinoless double beta decay depends on electron neutrino mass($m_{ee}$)~\cite{DellOro:2016tmg}, where
\bea
m_{ee} = \sum_i m_i |U_{ei}|^2\,.
\eea
Here, $U_{ei} $ is the elements of PMNS mixing matrix. 
The present status of neutrinoless double beta decay by these experiments is as follows:
The GERmanium Detector Array (GERDA) which uses germanium detector enriched in $^{76}{\rm Ge}$ achieved an upper limit
on $m_{ee} < 0.2$ eV in phase I of this experiment~\cite{Agostini:2013mzu}. In phase II of this 
experiments, GERDA achieved an improved upper limit of 0.15 eV~\cite{Agostini:2017iyd} on $m_{ee}$. 
Another experiment, Enriched Xenon Observatory (EXO) provides an upper limit, 
$m_{ee} < 0.19$ eV in the first run~\cite{Albert:2014awa}. The second phase of this experiments
is expected to bring down the upper limit to 0.09 eV~\cite{Maneschg:2017mzu}. 
The combined analysis of KamLAND-Zen-I and EXO-I puts an upper limit of $m_{ee} < 0.12$ eV~\cite{Gando:2012zm}. 
Second phase of KamLAND-Zen run achieved an upper limit of $m_{ee} < 0.06$ eV~\cite{KamLAND-Zen:2016pfg}. This is the best 
limit achieved so far. All these results mentioned above are at 90 $\%$ C.L.
We calculate the $m_{ee}$ using our $m_0$ and the other parameters
from oscillation data~\cite{Capozzi:2016rtj} in 2-$\sigma$ level too.
Our results shows that (see \fig{fig:double-beta}) some part of the allowed range of $m_0$ for normal 
hierarchy is already ruled out by KamLAND second run result, but for  inverted hierarchy allowed $m_0$
is still below the sensitivity of the experiment.
\section{Discussion and Conclusion}
\label{sec:conc}
The tensions between the results of LSS experiment and Planck CMB observation are well studied in the literature. 
These tensions were believed to be the signature of some unknown, interesting and exotic physics. 
However, due to relevance of the current investigation, we have discussed only two  such solutions 
namely effective viscosity on large scales and massive neutrinos throughout this paper. We have shown that the former one 
is a comparatively better solution to the problem than the later one. Nonetheless, from theoretical point of view
effective viscous theory is just a better description of dark matter fluid on large scale and it is something very
standard that should be naturally incorporated in the analysis. For simplicity, we have kept the viscosity coefficients
constant for our analyses. 
We would like to stress that time varying viscous coefficient won't change the conclusion drawn in the paper, but it 
might only change the required value of the viscosity parameter for resolving the tension between LSS and Planck data.

The amount of constant effective shear viscosity ($\eta$) required is $6\times 10^{-6}M_{P}^2 H_0$. This result is
of the same order in $1 h\,{\rm Mpc^{-1}}$ scale if we estimate the value of constant effective viscous coefficients
using the assumption made in the two loop calculation of matter power spectrum~\cite{Blas:2015tla} or the N-body simulation
result in ref~\cite{Carrasco:2012cv}. Therefore we 
can safely claim that this viscosity required for resolving LSS-Planck tension is of effective in nature.

Therefore, under this improved version of dark matter fluid on large scales, the scopes of introducing exotic physics
in cosmological scenario is reduced. LSS-Planck tension has been advocated as a signature of massive neutrinos in 
cosmology many times in past~\cite{Wyman:2013lza,Battye:2013xqa}. But in the effective viscous description of dark matter fluid,
provision of accommodating massive neutrinos becomes more constrained. 
Moreover, the lower bound on lightest neutrino mass again touches zero value, making existence of one massless 
neutrino absolutely viable.

This approach of comparing different proposed solutions to some particular cosmological observations and then comparing
them to constrain one using another is the central theme of this paper. We expect in future other exotic physics, like
dark-matter neutrino interaction, presence of dark radiation, two component dark matter, dark energy-neutrino interaction etc.,
will be constrained in a better way using this approach.

\bibliographystyle{JHEP}
\bibliography{paper_draft.bib}  
  
\end{document}